\documentclass[final,authoryear,5p,times]{elsarticle}

\usepackage{amssymb}
\usepackage{amsthm}
\usepackage{longtable}
\usepackage{url}
\usepackage{natbib}
\usepackage{amsmath}
\usepackage{graphicx}
\usepackage{longtable}
\usepackage{lastpage}
\def\eg{{\it e.g.~}}

\def\ie{{\it i.e.~}}
\def\km2{\ensuremath{\mathrm{km}^{2}}}

\def\SAf{\ensuremath{\Sigma Af}}

\def\Afrho{\ensuremath{Af\rho}}
\def\Bc{\ensuremath{Bc}}
\def\Rc{\ensuremath{Rc}}
\def\CN{\ensuremath{CN}}
\def\Cd{\ensuremath{C_\mathrm{2}}}
\def\Ct{\ensuremath{C_\mathrm{3}}}

\def\deg{$\,^{\circ}$}
\journal{Icarus}

\begin{document}

\begin{frontmatter}

\title{Activity of comet 103P/Hartley 2 at the time of the
  EPOXI mission fly-by\thanks{
Based on Observations performed at the European
    Southern Observatory, La Silla, Progr. 086.C-0375. This work has
    been partially funded by MIUR - Ministero dell'Istruzione,
    dell'universit\'a e della Ricerca, under the PRIN 2008 funding.  }
}
\author[OAA]{Gian Paolo Tozzi
}
\author[OAN]{Elena Mazzotta Epifani}
\author[ESO]{Olivier R. Hainaut}
\author[OAA]{Patrizio Patriarchi}
\author[IAA]{Luisa Lara}
\author[OAA]{John Robert Brucato}
\author[MPS]{Hermann Boehnhardt}
\author[OCA]{Marco Del B\'o}
\author[IAC]{Javier Licandro}
\author[IFA]{Karen Meech}
\author[OCA]{Paolo Tanga}

\address[OAA]{INAF -- Osservatorio Astrofisico di Arcetri, Largo E. Fermi 5,
I-50\,125 Firenze, Italy}
\address[OAN]{INAF -- Osservatorio Astronomico di Capodimonte, Via
  Moiariello 16, I-80\,131 Napoli, Italy}
\address[ESO]{European Southern Observatory --
  Karl-Schwarzschild-Stra\ss e 2, D-85\,748 Garching bei München, Germany}
\address[IAA]{Instituto de Astrofis\'{\i}ca de Andaluc\'{\i}a (IAA-CSIC)
C/ Glorieta de la Astronom\'{\i}a,s/n
18008 Granada, Spain
}
\address[MPS]{Max-Planck Institut f\"ur Sonnensystemforschung, D-37\,191
Katlenburg-Lindau, Germany}
\address[OCA]{UNS-CNRS-Observatoire de la C\^ote d'Azur,
Laboratoire Cassiop\'ee, BP 4229, 06\,304 Nice cedex 04, France}
\address[IAC]{Instituto de Astrof\'{\i}sica de Canarias, V\`{\i}a L\`actea s/n,
38\,200 La
Laguna, Tenerife, Spain}
\address[IFA]{Institute for Astronomy -- University of Hawai`i, 2680
  Woodlawn Drive, Honolulu, HI 96\,822, USA}

\begin{abstract} 

Comet 103P/Hartley~2 was observed on Nov. 1-6, 2010, coinciding with the fly-by
of the space probe EPOXI. The goal was to connect the large scale
phenomena observed from the ground, with those at small scale observed
from the spacecraft. The comet showed  strong activity correlated with
the rotation of its nucleus, also observed by the spacecraft.  We
report here the characterization of the solid component produced by this
activity, via observations of 
the emission in two spectral regions where only grain scattering of
the solar radiation is
present. We show that the 
grains produced by this  activity 
had a lifetime of the order of 5 hours, compatible with the spacecraft
observations of the large icy chunks. Moreover, the grains
produced by one of the active regions have a very red color. This suggests an
organic component mixed with the ice in the grains.
\end{abstract}

\begin{keyword}
Comets \sep Comets, dust \sep Comets, coma
\end{keyword}
\end{frontmatter}

\section{Introduction}
\label{intro}

Comet 103P/Hartley 2 (hereafter 103P) was discovered in March 1986 by
M.~\citet{Hartley1986} with the UK Schmidt Telescope at Siding
Spring (Australia). Its dynamical history \citep[][and
following electronic updates]{Carusi1985} shows that its orbit has been quite
unstable over the last 150 years, with a perihelion distance
oscillating between 1 and 2.5~AU. 103P is one of the few comets that
have become an Earth crosser in the recent past. 


103P has been frequently observed over the 20 years following its
discovery, both by ground-based and space telescopes. The portrait
that emerged from this harvest of ground-based data
\cite[\eg][]{Licandro2000, Lowry2001, Lowry2003, Snodgrass2006,
  Snodgrass2008, EME2008} is that of a highly active comet, even at
large heliocentric distance \citep[5~AU,][]{Snodgrass2008}. In
October 2007, 103P was selected as the target for the NASA Deep Impact
extended mission EPOXI \citep{AHearn2005}; consequently, an intense
world-wide observation campaign has been devoted to characterize its
nucleus and coma properties in order to prepare for the spacecraft
fly-by, occurring in November 2010.

The main results of this campaign are summarized in \citet{Meech2011}:
the comet has a small, sub-km, nucleus, with a rotation period of
16.4~hrs when inactive, slowly increasing with activity. This possibly
indicates that the rotation rate is slowed by out-gassing from the
(irregular) surface.

The campaign data also showed that the active fraction of the nucleus'
surface is, as typical, about 2\%, but that it is surrounded by a large
halo of (icy?) grains that contribute more than the nucleus to the
total water production rate ($>$ 90\% at perihelion). The presence of
large grains was already inferred during the 1998 perihelion passage:
analysis and modeling of ISOCAM \citep{Cesarsky1996} infrared images of
the dust coma and tail  \citep{Epifani2001} implied an evolution
of the dust production rate from 10~kg~s$^{-1}$ at 3.25~AU to
100~kg~s$^{-1}$ at 1.04~AU, with grains up to centimeters in
size. This dust environment of 103P seems consistent with a {\it
  trail} structure \citep{Lisse2009}, presumably associated with
millimeter-sized debris.


On UT 4.583 November 2010, the NASA mission spacecraft EPOXI flew by 103P. The
closest approach was 694~km, when the comet was at 1.064~AU from
the Sun. The main results of the {\it in-situ} measurements are
described in \citet{AHearn2011}: the nucleus showed a
bi-lobed morphology, with a maximum length of 2.33~km and a mean
radius of $0.58\pm 0.02$~km. The rotation period at the time of the
closest approach was measured to be $18.34\pm 0.04$~h. Images obtained
during the fly-by confirmed the presence of individual, ``large''
chunks near the nucleus, moving at 1--2~m~s$^{-1}$. Large grains had
already been detected via radar observations just before the close
encounter \citep{Harmon2011}: decimeter-sized grains (or possibly even
larger), moving at 20--30~m~s$^{-1}$, and ejected into free
trajectories rather than circum-nuclear orbits were modeled to fit
the grain-coma echo from 103P. \citet{AHearn2011} argued that the
largest chunks they detected from EPOXI were icy, with radii up to 10--20~cm,
dragged out by super-volatiles (specifically, CO$_2$) and then
sublimating to provide a large fraction of the total H$_2$O gaseous
output of the comet.

Here we report observations done during 5 (half) nights around the
time of the space probe fly-by. The observations
were obtained with narrow band filters centered in regions with continuum
emission, \ie\ due to the scattering of solar radiation by the
grains present in the coma.

\section{Observations and data reduction}

All the observations were performed with
EFOSC\footnote{http://www.eso.org/sci/facilities/lasilla/instruments/efosc/}
at the ESO 3.56~m New Technology
Telescope (NTT), in La Silla (Chile).
The observation epoch, geometry and conditions
are listed in Table~\ref{tab_log}.

\begin{table*}
\begin{center}
\caption{ \label{tab_log} Log of Observations through narrow band continuum
cometary filters in the visible.}
\begin{tabular}{cccccccccc}
\hline
Date  &UT$_{str}$&  r$_h$ &  $\Delta$ & Phase & PA & Sky & Filt. & T$_{exp}$
& airm \\ 
2010 Nov.& hh:mm    &  AU    & AU        &\deg &\deg&       &      & s         &  
  \\ 
\hline
\hline
01    & 07:59    & 1.06   & 0.143  & 58.4  &280.4 & Pht    & \Bc& 5$\times$60 &
1.43\\ 
"     & 08:18    & "      &  "     & "     & "    &  "    & \Rc& 5$\times$60 & 1.36\\ 
"     &08:25     & "      & "      & "     & "    &   "   & \Bc& 5$\times$120&
1.34 \\ 
"     &08:57     & "      &  "     & "     & "    &   "   & \Rc& 7$\times$120& 1.30 \\ 
03    &08:10     & 01:06  &  0.151 & 58.7  & 282.6&   Pht  & \Bc&
5$\times$180&1.30  \\ 
"     &08:54     & "      &  "     & "     & "    &   "   & \Rc& 5$\times$180&1.27  \\ 
04    &07:43     & 1.06   & 0.154  & 58.8  & 283.7& Clr   & \Bc& 5$\times$180&1.31  \\ 
"     &08:04     & "      &  "     & "     & "    &   "   & \Rc& 5$\times$180&1.27  \\ 
"     &08:48     & "      &  "     & "     & "    &   "   & \Bc& 5$\times$180&1.24  \\ 
05    &05:10     & 1.06   & 0.156  & 58.8  & 284.6&   Pht  & \Bc&
8$\times$120&2.30  \\ 
"     &05:34     & "      &  "     & "     & "    &   "   & \Bc& 8$\times$120&1.97  \\ 
"     &05:58     & "      &  "     & "     & "    &   "   & \Rc& 8$\times$120&1.74  \\ 
"     &06:45     & "      &  "     & "     & "    &   "   & \Bc& 8$\times$120&1.46  \\ 
"     &07:08     & "      &  "     & "     & "    &   "   & \Rc& 8$\times$120&1.37  \\ 
"     &07:56     & "      &  "     & "     & "    &   "   & \Bc& 8$\times$120&1.26  \\ 
"     &08:19     & "      &  "     & "     & "    &   "   & \Rc& 8$\times$120&1.23  \\ 
"     &08:46     & "      &  "     & "     & "    &   "   & \Bc& 8$\times$120&1.21  \\ 
06    &05:03     & 1.06   & 0.163  & 58.8  &285.6 &  Pht   & \Rc&
8$\times$120&2.32  \\ 
"     &05:27     & "      &  "     & "     & "    &   "   & \Rc& 8$\times$120&1.97  \\ 
"     &05:50     & "      &  "     & "     & "    &   "   & \Bc& 8$\times$120&1.75  \\ 
"     &06:38     & "      &  "     & "     & "    &   "   & \Bc& 8$\times$120&1.46  \\ 
"     &07:01     & "      &  "     & "     & "    &   "   & \Rc& 8$\times$120&1.36  \\ 
"     &07:48     & "      &  "     & "     & "    &   "   & \Bc& 8$\times$120&1.25  \\ 
"     &08:12     & "      &  "     & "     & "    &   "   & \Rc& 8$\times$120&1.21  \\ 
\hline
\end{tabular}
\end{center}
UT$_{str}$ refers to the beginning of the observations; $r_h$ and
$\Delta$ are the helio- and geocentric distances; {\it Phase} is the
Sun-Target-Observer angle; {\it PA} is the position angle of the extended
Sun-Target vector. The sky conditions are listed: {\it Pht} is for photometric,
{\it Clr} is for clear. {\it Filt.} is the identifier of the filter;
{\it T$_{exp}$} the
is the sequence exposure time on target, in second; the airmass ({\it airm}) is
listed for
the beginning of each observations
\end{table*}

Most of the observations consisted in images of the comet obtained
through Narrow Band (NB) cometary filters \cite[similar to those described
  in][]{FSA00}.  The NB filters included blue and red continuum (\Bc\ 
and \Rc), \CN (0-0) Violet band, \Ct\ $^1\Pi_u - ^1\Sigma_g^+$ and the
\Cd\ Swan ($\Delta\nu=0$) bands. In this paper we focus on the solid
component of the coma, \ie\ the data obtained with the two continuum
filters.  Table \ref{NB_filt} lists their central wavelengths and full
widths at half maximum.

To minimize the contamination by background stars and to evaluate the
sky background, each comet observation sequence consisted of 5 or 8
exposures on target, moving the telescope by few tens of arcseconds in
between, and one additional exposure obtained $\simeq$ 8$'$ off the
comet, to record the sky uncontaminated by the comet.

Bias and twilight sky flat-field exposures were also obtained in order
to correct for the instrumental signature.  Spectrophotometric
standard stars and solar analog stars were observed spectroscopically 
and also in imaging mode, with the same
filters at about the same airmass as the comet, to
calibrate in flux the comet images.


At the beginning of the run, some images of the comet were acquired  through the
standard broad band filters $V$, $R$ and $i$. They
were  not used in this analysis, because their pass-bands contain non
negligible gas emission lines. This was verified \textit{a posteriori}
with the spectra.

\begin{table}
\caption{\label{NB_filt}Narrow band continuum cometary filters:
  central wavelength and FWHM}
\begin{center}
\begin{tabular}{ccc}
\hline
Filter Name & Central $\lambda$ & FWHM \\
            & [\AA]             & [\AA]\\
\hline\hline
\Bc & 4430 & 33\\
\Rc & 6840 & 74\\
\hline
\end{tabular}
\end{center}
\end{table}


All the images and spectra were at first corrected for bias and
flat-field, using the appropriate ancillary frames in the customary
manner. To calibrate the images in $Af$ (see below) we computed the
``theoretical'' filter color indexes (\Bc-V and \Rc-V) of the observed
spectrophotometric standard and solar analog stars.  By using the
tabulated spectra, we compared stellar fluxes measured through the NB and
V
filters  to that of a star of A0V spectral type that, by
definition, has a color index equal to 0. The knowledge of the V
magnitude of the observed stars allowed us to recover the NB
magnitudes and hence, from the observations, the photometric zero point (ZP) of
each NB
filter for each night.

For the extinction correction, we adopted the standard extinction of
La Silla\footnote{
  http://www.ls.eso.org/sci/facilities/lasilla/astclim/atm\_ext/}.
Since the standard stars were observed at about the same airmass as at
least one of the comet sequences, the errors introduced by possible
differences in the extinctions are negligible. When more than one
sequence with the same filter was observed, the resulting calibrated
images were in agreement within 10\%. The same agreement was found
also for calibrated images from consecutive photometric nights, in
regions of the coma where the signature produced by periodic change of
activity (see below) was not yet present.  All the frames with the
same filter were then inter-calibrated (see below).

The level of the sky contribution was then evaluated for each comet sequence,
measuring the median level of the sky frame acquired
through the same filter $8'$ away from the comet. This value was
subtracted from each frame.

The sky-subtracted frames then were re-centered on the comet
photometric center, and a composite comet image was obtained through a median
average of the 5 or 8 frames of the sequence.

The use of a median combination significantly reduces the contamination
produced by background stars.  The background was then refined and
subtracted with a trial and error procedure using the \SAf\ function (see
below), by making
this function independent of the projected
nucleo-centric distance ($\rho$) for distances greater than 150 pixels 
(corresponding to $\simeq$ 15000 km at the comet distance)
from the photometric center.  The resulting continuum images were then
calibrated in $Af$ \citep{AHearn1984}.

\section{Analysis}

\subsection{1D analysis: background activity}

As shown by the spacecraft observations \citep{AHearn2011} and as already
noticed at the telescope during the observations, the comet showed a
strong variability with nucleus rotation. By studying the
CN features in the coma, \citet{Samarasinha2011} found that rotation
period varied from 17.1~h in September to 18.8~h in November 2010. Thus for
this study we assume a rotation period of 18.8~h. We arbitrarily use as a
starting
point for the rotation phase the time of the first observation (not used
here because it was recorded with a broad band filter).
This  point was Nov. 1, UT = 7:39.

To check how the emission of the continuum varied with the rotation
period, we first characterized the constant background coma, which was
estimated from the epoch of minimal activity.

\citet{AHearn1984} introduced the function \Afrho\ as a proxy for production of
the
solid component. $A$ is the geometric albedo of the grains and $f$
the filling factor, defined as the percentage of the  area
that is covered by the grains, and $\rho$ the projected nucleo-centric
distance of the aperture. Typically, the filling factor, $f$, is
proportional to $1/\rho$, while $A$ is, at first approximation,
independent of $\rho$. Since from the observations we get the product
$Af(\rho)$, it is not possible to
disentangle these two parameters without additional observations in
the thermal IR or without making assumptions.

Here we used another function derived by the above one, \ie 
\SAf($\rho$), that
is proportional
to the average column density of the solid component at the projected
nucleo-centric distance $\rho$. It is equal to $2\pi \rho Af(\rho)$. As
shown by \citet{Tozzi2007}, \SAf\ should be constant with respect to the
projected nuclear distance, $\rho$, for a comet with a dust outflow of
constant velocity and production rate, and if sublimation or
fragmentation of the grains are excluded. The solar radiation pressure
introduces a small linear dependence with $\rho$, but normally its
effects are only noticeable at large distances from the nucleus, larger than
the field of view (FoV) of EFOSC.

\begin{figure} 
\centering
{\bf a.}\includegraphics[width=8.5cm]{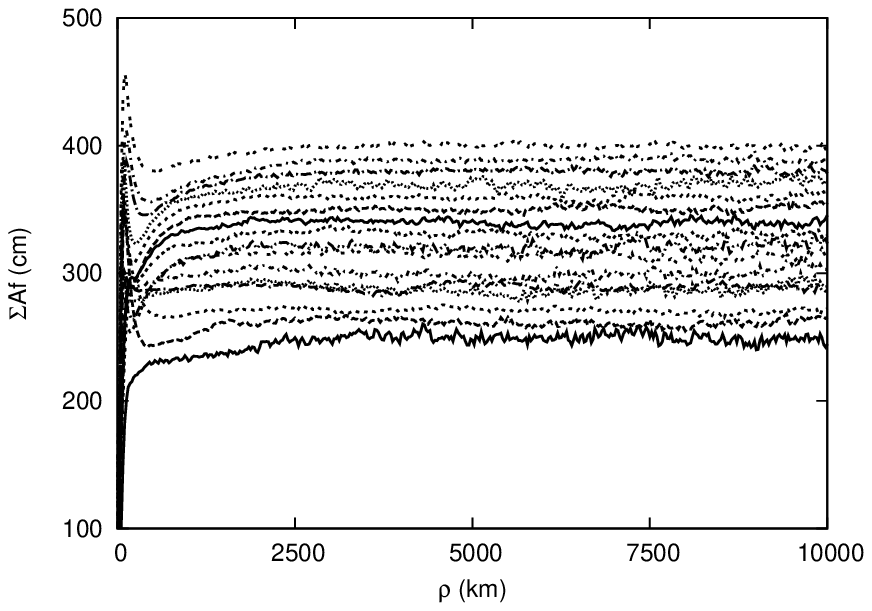}\\
{\bf b.}\includegraphics[width=8.5cm]{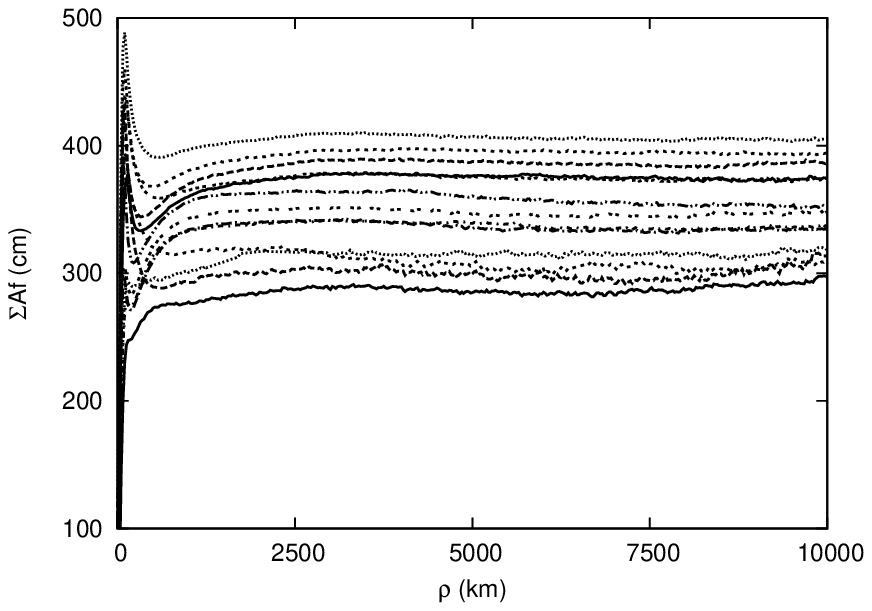}

\caption{\SAf($\rho$) functions for all the observations obtained with
  \Bc\ (a) and \Rc\ (b) filters. The functions have been vertically
  shifted by 10~cm for clarity. The plots are ordered (from bottom to
  top) in observational sequence, as in Table~\ref{tab_log} }
\label{fig_SAf}
\end{figure}

The final calibrated images were analyzed by computing their
\SAf($\rho$) function.  In Fig.~\ref{fig_SAf} some examples of the
\SAf\ function are shown.  The solid component production rate was
clearly not constant, as it can be noticed by comparing the regions
with $\rho <$ 500-1000 km.  However, excluding the region with $\rho$
smaller than about 3000--5000~km, the profiles are very similar, and
the signatures of the change of activity seems to overlap that of a
typical, constant profile.

Plotting all the \SAf\ profiles together, as shown in
Fig.~\ref{fig_SAf}, we can see that they have the same behavior for
$\rho$ greater that 3000-5000 km and the values of \SAf\ in that
region are within $\pm 10 \%$.  These differences are comparable to
the uncertainty of the absolute calibration.  We therefore assumed
that the average value of \SAf\ over $\rho$ in the 6000--8000~km range
was constant over the observations, and made small corrections to the
Zero Points to adjust the profiles so they have the same average value
over that range.  This assumption is equivalent to considering that
any change in grains production did not reach $\rho = 6000$~km from
one day to the next.  The validity of this hypothesis will be verified
later. Note that using this method, the data acquired during the
non-photometric night (Nov.~4) have been also calibrated to the same
system as the others.

\begin{figure} 
\centering
\includegraphics[width=8.5cm]{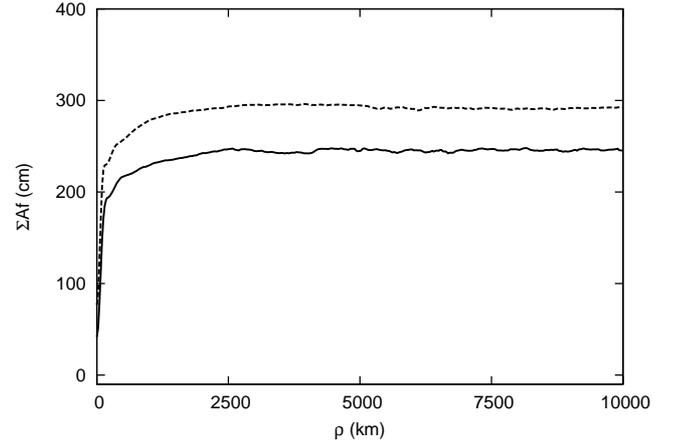}

\caption{\SAf\ profiles  corresponding to the comet at minimum
activity. The solid line is for filter $Bc$, and the dashed line for $Rc$.
\label{fig_SAf_min}}
\end{figure}

All the \SAf\ profiles are very similar except in the region very
close to the nucleus ($\rho <$ 2000 km) where the signature of the
activity seems confined. In order to characterize the periodic
emissions, a \SAf\ profile corresponding to the minimum of cometary
activity --what we call the ``quiet comet''-- was determined for each
filter as
follows: for the regions with $\rho >$ 6000 km  as a
median of all
the profiles, and for that with
$\rho < $6000 km as the minimum envelope of all the profiles (excluding the
region
with $\rho< 200$ km). The  median was used to reduce 
signatures of possible background stars, still present in the single \SAf\
profiles; the minimum envelope allowed us to discard the peaks produced by the
activity. However, the always present activity in the regions with $\rho <
$100-200  km (see below) was visually removed by spline interpolation. The
\SAf\ 
profiles, 
corresponding to the
constant level of activity, are shown in Fig.~\ref{fig_SAf_min} for
both filters.

From the measurements of \SAf, the values of \Afrho\ for $\rho$ in the
6000--8000~km range corresponded to 80$\pm$3~cm and 91$\pm$3~cm in
\Bc\ and \Rc, respectively. The errors were obtained from the standard
deviation of the values of the \SAf\ function, with the
re-normalization described above.

From the above values of \Afrho, it is possible to derive the slope of
the reflectivity spectrum for the solid component coma of the ``quiet
comet''.  Assuming a linear variation with lambda, the \Afrho\ at
$\lambda$=5550~\AA\ would be 85$\pm$3~cm and the corresponding spectral slope
would be
5.2$\pm$1.5~\%/1000~\AA.

As seen in Fig.~\ref{fig_SAf_min}, the \SAf\ profiles of the minimum
activity are not completely constant with $\rho$, but they
systematically increase in the inner region of the coma up to 
$\rho$ equal to 2000--2500~km. Note that this behavior cannot be due to a
residual background, because this would produce a {\it linear} variation with
$\rho$.  A profile like that means that the total grain
cross-section increases with $\rho$ in a systematic way. The only possible
explanation is the fragmentation of large grains, with dimensions much
larger than the observation wavelength, as they move away from the
nucleus. In that way, the total grain cross-sections would increase
with $\rho$. It is important to notice that the behavior of the quiet coma can
be found in all the profiles, even though,
it is partially hidden by the periodic activity


\subsection{Periodic activity}

\begin{figure} 
\centering
\includegraphics[width=8.5cm]{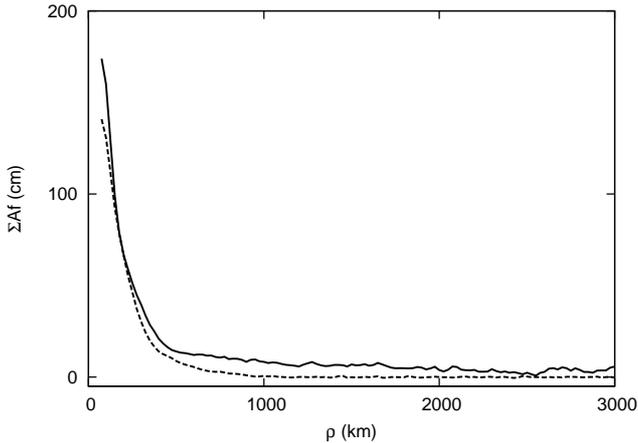}

\caption{Example of \SAf\ profiles produced by the periodic activity:
  the profile corresponding to the minimum activity was subtracted
  from an individual profile, leaving only the profile for additional
  activity.  The solid line is for the observations through the \Bc\ filter,
the dashed
  line for observations through the \Rc\ filter.  }
\label{fig_SAf_burst}
\end{figure}

By subtracting the minimum profile for \Bc\ or \Rc\ from the individual
\SAf\ profiles, we found the signature of the clouds of grains
periodically ejected by the nucleus. Typical \SAf\ profiles of the
 those clouds are shown in Figure~\ref{fig_SAf_burst}. All profiles are
similar, with a very strong increase towards the nucleus. There
is no evidence of any motion of the clouds, as it has been seen for
other outbursts \cite[see for instance][]{Tozzilic2002}.

It would be interesting to determine the colors of the clouds of grains
produced by the periodic activity,
and to compare them to that of the coma at minimum activity. This
measurement is complicated by the rapid evolution of the cloud and by
the fact that the observations in \Bc\ and \Rc\ are taken at different
times.  To measure the color of the material produced by the activity, we
have first to
characterize its evolution with time.

We determined then the time when the comet was closer to the minimum
activity, by checking the individual \SAf\ profiles and then selecting the
minimum ones. They were those recorded on November 6 at UT 5:10 and
5:58 for the \Bc\ and \Rc\ filters, respectively.  Their \SAf\ profiles are very
similar to the minimum profiles derived above: only a narrow and faint peak is
present in the region with $\rho <$ 100-200 km, indicating that the
periodic activity was starting again. By chance those two observations
are very close to the rotation phase equal to 0 as defined in the
previous section. 

To map the emission produced by the periodic activity, we subtract these
images corresponding to the minimum of activity from the individual images.

As pointed out above, those two images were not acquired at the exact 
minimum of activity: the inner part already shows the signs of the
periodic activity emission.  Nevertheless, these signs are limited to
the very inner coma ($\rho \lesssim$ 200 km) and it is easy to take
them into account in the following analysis.

\begin{figure*} 
\centering
\includegraphics[width=16cm]{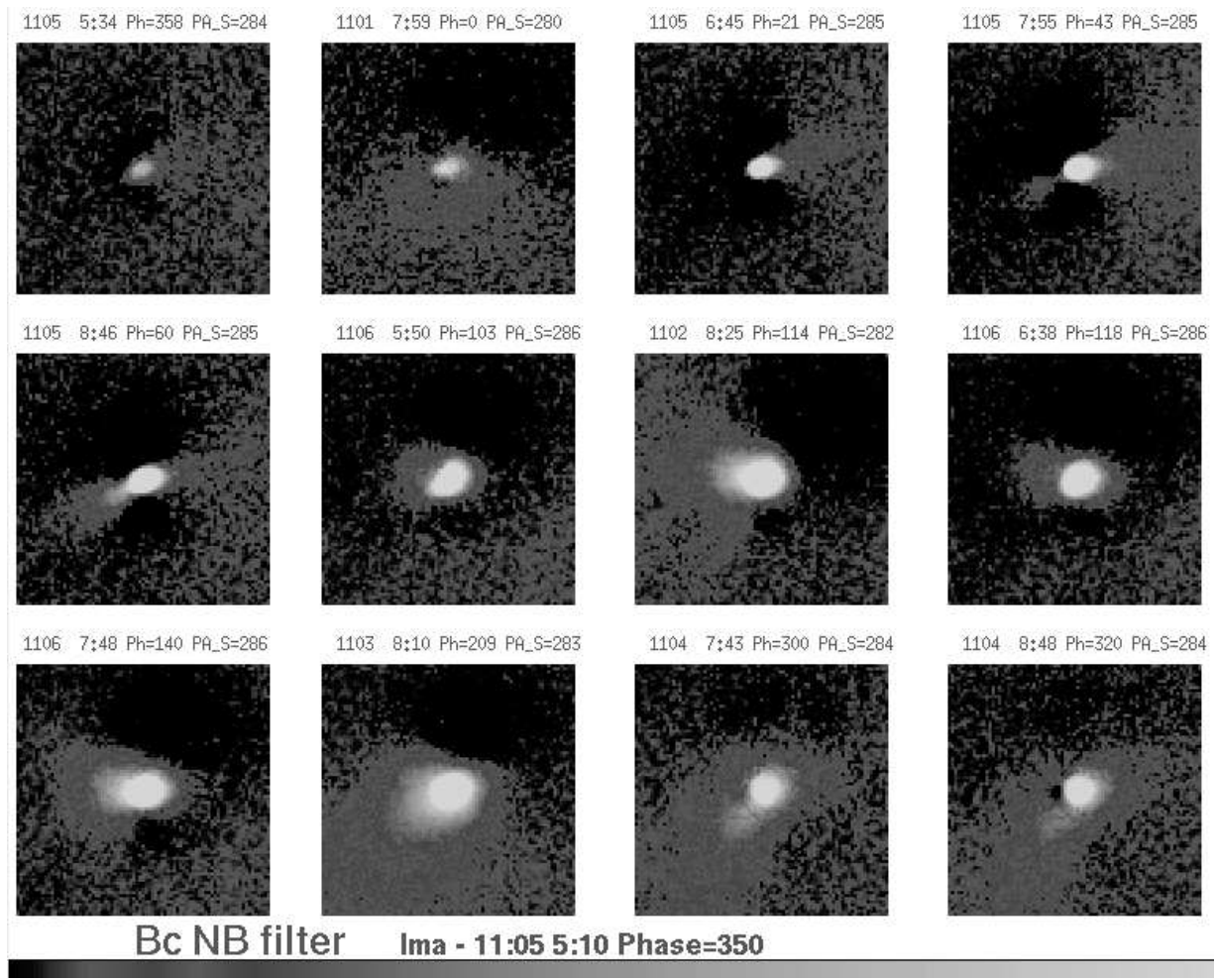}
\caption{Maps of the material released by the periodic activity of the
comet
as observed with the  \Bc\ filter. The images are differences of the
comet images minus that of Nov. 5,
5:10 UT, the one with the minimum periodic activity. Each sub-image covers a
projected area FoV
of 2000$\times$2000 km$^2$, with the comet at the center. North is up, East to
the left. The
sub-images have been ordered with the rotation phase (marked Ph). PA\_S is the
position angle of the
projected Sun direction. The look-up table is logarithmic.} 
\label{fig_Bc_ejecta}
\end{figure*}

\begin{figure*} 
\centering
\includegraphics[width=16cm]{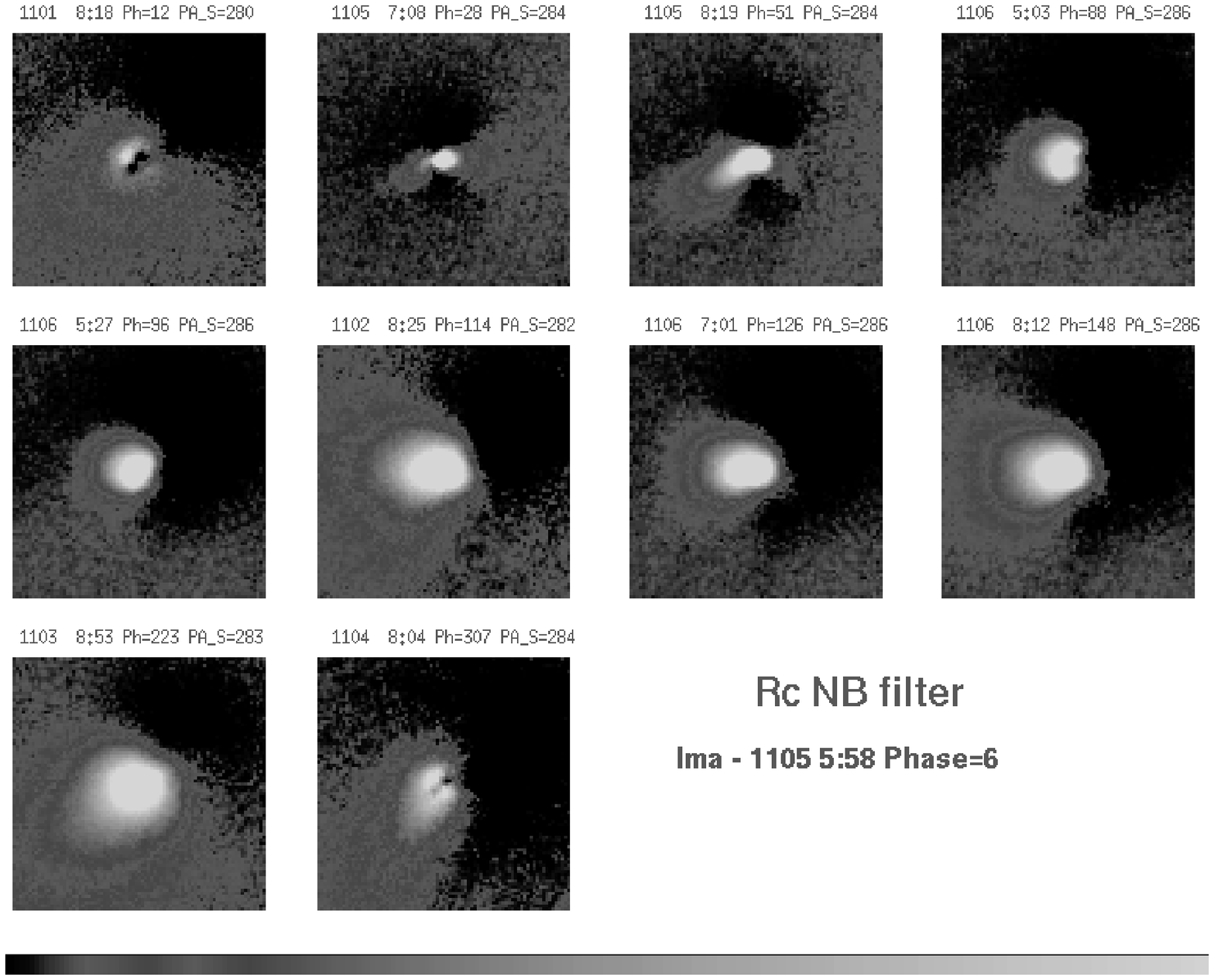}
\caption{Maps of the material released by the periodic activity of the
  comet as observed with the  \Rc\ filter. The images are
  differences of the comet images minus that of Nov. 5, 5:58 UT, the one
with minimum periodic  activity. Each sub-image covers a projected area FoV of
  2000$\times$2000 km$^2$, with the comet at the
  center. North is up, East to the left. The sub-images have been ordered with
the rotation
  phase (marked Ph). PA\_S is the position angle of the
projected Sun direction. The look-up table is logarithmic.  }
\label{fig_Rc_ejecta}
\end{figure*}

The activity maps are shown as a function of the rotation phase in
Figures~\ref{fig_Bc_ejecta} and \ref{fig_Rc_ejecta} for the \Bc\ and
\Rc\ filters, respectively.  For clarity, the figures
cover a limited FoV, equivalent to a projected area of
2000 $\times$ 2000~km$^2$ centered on the comet. The FoV actually
covered by the observations is more than 10 times larger.
The grains released by
the periodic activity are clearly visible in both
filters. They are particularly evident in two preferred
directions. The first one points to the East at position angle PA$\sim$90\deg
and is active from a rotation phase of about 90\deg\ to
200\deg. The second one points to PA $\sim$140\deg and is active
from the rotation phase greater than 200\deg. 
They don't seem produced by the same active region on the rotating nucleus,
because in such a case we should see the produced grains spiraling around the
nucleus.

\begin{figure} 
\centering
\includegraphics[width=8.5cm]{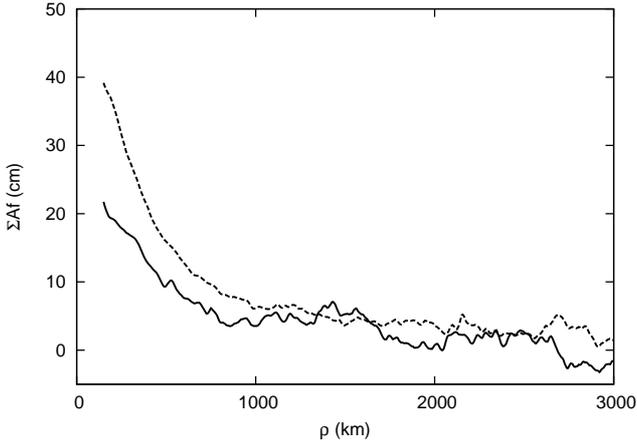}

\caption{Example of equivalent \SAf\ profiles produced by the cloud of
particles released by the activity at
PA$\sim$90\deg:
  Solid and dashed
  lines are for observations in \Bc\ and \Rc, respectively.  }
\label{fig_SAf_PA080}
\end{figure}

We have then analyzed the two directions independently, to check
whether the active regions have different origins, as found by the
spacecraft \citep{AHearn2011}.  The maps have been transformed to
polar coordinates, and the profiles of the emission with respect to
the nucleo-centric distance $\rho$ have been obtained by integrating
between PA~=~40\deg\ and 117\deg\ for the first cloud, and between
PA~=~118\deg\ to 165\deg\ for the second one. The profiles have been
transformed in equivalent \SAf\ units (accounting for the limited
angular range, as the standard definition of \SAf\ implies an
integration over 2$\pi$).  Typical equivalent \SAf\ profiles of the
cloud during a phase of high activity are shown in
Fig.~\ref{fig_SAf_PA080}. 

The \SAf\ profiles for the active regions are very similar to those
presented earlier for the 1D analysis --only the peak intensity and
signal-to-noise ratio (SNR) are different.  Excluding the region with
$\rho \lesssim$ 200~km, which is dominated by the seeing of the image
and contaminated by the near-nucleus activity in the reference image
of the ``quiet'' comet, the profiles are very well represented by a
constant plus an exponential function,
$$\SAf(\rho) = \SAf_0 + \SAf_1 \exp{(-\frac{\rho}{L_1})},$$ where
$L_1$ is the scale length of the cloud released by the activity (in km),
and $\SAf_1$ the peak
intensity (in cm). The fit residuals are very small, with errors less than
10\% even for
profiles with medium/high activity, \ie\ with high SNR. 

The variation of $L_1$ with time gives the projected expansion
velocity of the grains located at $\rho = L_1$. Of
course grains located at $\rho > L_1$ have larger projected velocities
and those at smaller distances have smaller velocities. Since
the
direction of the emission of the cloud of particles in space is not known,
the projected velocities
are a lower limit of the real expansion velocities.  To get the projected
velocity we used the
observations from November~6, that are most numerous and that were obtained
when the level of activity was relatively high. The resulting ${dL_1}/{dt}$ is
$\approx 15$--20~m~s$^{-1}$ for both clouds, which is comparable to the
velocity
of the grains measured by radar \citep{Harmon2011}. Note that they measured
radial velocities, while we measured projected velocities.

Those low projected velocities imply that the cloud of grains produced by
periodic activity cannot move to
distances greater than about 1700 km in 24 hours. This implies
the grains must still be well within the FoV for observations recorded
during the following night. The assumption we had made before, that
the grains had not reached a projected distance larger than 6000 km
from one night to the next, is therefore valid, justifying our
inter-normalization of the outer part of the profiles.

The equivalent $\SAf(\rho)$ functions are always
peaked at $\rho$ close to 0, as can be seen in Fig.~\ref{fig_SAf_burst}. A
motion of the cloud away from the nucleus was never detected (apart the
small
increase of $L_1$ mentioned above)
contrary to other comets, \eg C/1999~S4 before its breakup
\citep{Tozzilic2002} or 9P/Tempel~1 after the impact
\citep{Tozzi2007}, for which a similar analysis revealed a clear motion of
the ejecta with a Maxwellian profile.

The product $\SAf_1 \times {L_1}$ gives the cross-sections, $SA$, of the
released clouds, \ie\ the total surface ($S$) covered by the grains
multiplied by their geometric albedo ($A$). 

\begin{figure} 
\centering
{\bf a.}\includegraphics[width=8.5cm]{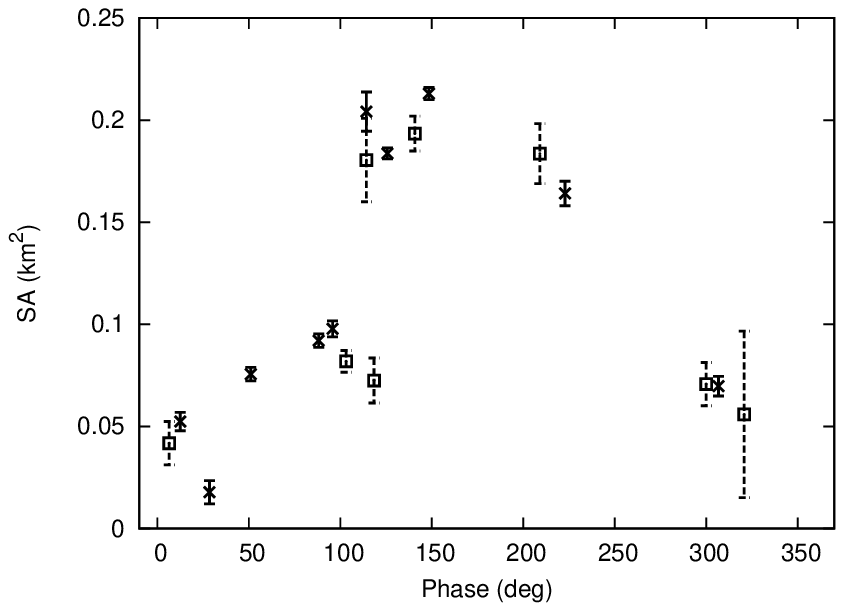}\\
{\bf b.}\includegraphics[width=8.5cm]{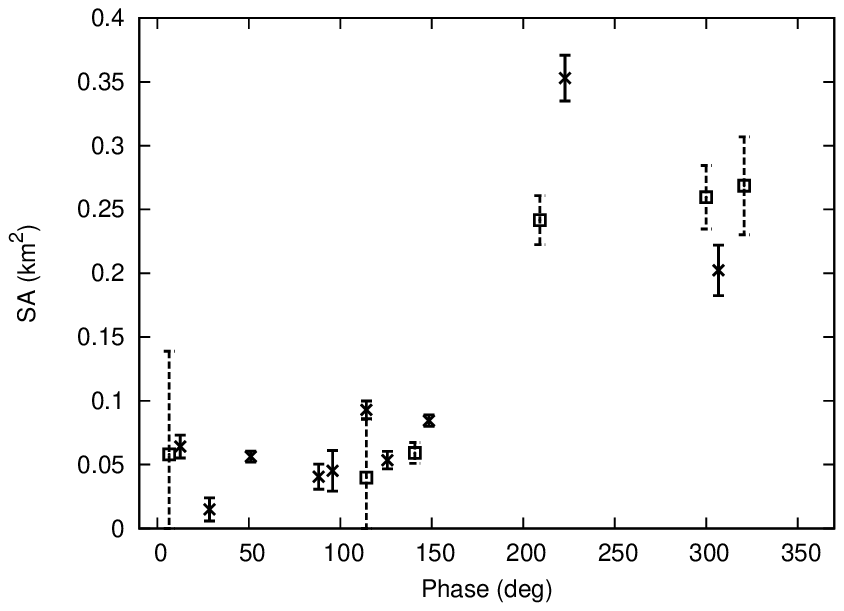}

\caption{Cross-section $SA$ in km$^2$ of the cloud emitted toward  East (a.)
and
  South-East directions (b.), as a function of the nucleus rotation phase.
Squares are derived from observations through the \Bc\ filter, and crossed
through the \Rc\ filter. For the cloud released into the East direction, 
  the \Bc\ cross-sections have been multiplied by 1.36 (see text). All
  the other values are plotted without multiplications. Note that the error
bars are 3 times the value given in Table \ref{Bc_ejecta}
}
\label{fig_SA_phase}
\end{figure}

The values of $SA$
together with those of the
scale-length are reported in Table~\ref{Bc_ejecta} for both
filters. $SA$ as function of the rotation phase is shown in
Fig.~\ref{fig_SA_phase}.

\begin{table*}
\caption{\label{Bc_ejecta} Fitting parameters for the profiles of the
clouds produced by the periodic activity. {\it L} is the scalelength and {\it
SA} is the cross section, as explained in the text. }
\begin{center}

\begin{tabular}{ccccccc}
\hline
Date & UT & Rot.  & \multicolumn{2}{c}{P$\sim$A80\deg} & 
\multicolumn{2}{c}{PA$\sim$140\deg} \\
2010 Nov.& & Phase & $L$ & $SA$ & $L$ & $SA$\\
 & [h:mm] & [\deg] & [km] & [km$^2$]& [km] & [km$^2]$\\
\hline\hline
\multicolumn{3}{c}{\Bc\ filter}\\
\hline
01 &  7:59 &    6  & 385$\pm$26&  0.0307$\pm$0.0026 & 274$\pm$51  & 0.0427$\pm$0.0198 \\
02 &  8:25 &  114  & 452$\pm$14&  0.1327$\pm$0.0050&  100$\pm$14  & 0.0293$\pm$0.0112 \\
03 &  8:10 &  209 & 487$\pm$11&  0.1350$\pm$0.0036& 1272$\pm$30 & 0.1777$\pm$0.0047 \\
04 &  7:43 &  300 & 642$\pm$29&  0.0520$\pm$0.0026& 1422$\pm$40 & 0.1909$\pm$0.0061 \\
04 &  8:48 &  320 & 197$\pm$22&  0.0411$\pm$0.0100& 1946$\pm$88  &0.1975$\pm$0.0094 \\
05 &  5:34 &  358 &  -&  -&     -            &  - \\
05 &  6:45 &   21 &  - &  - &     -            &  - \\
05 &  7:55 &   43 &  - &  - &     -            &  - \\
05 &  8:46  &  59 &  -&  -& 281$\pm$13  & 0.0464$\pm$0.0038 \\
06 &  5:50 &  103 & 277$\pm$ 5&  0.0602$\pm$0.0013& - & - \\
06 &  6:38  & 118 & 264$\pm$11&  0.0533$\pm$0.0027& - & - \\
06 &  7:48 &  140 & 424$\pm$ 5&  0.1422$\pm$0.0021& 216$\pm$6 & 0.0435$\pm$0.0020 \\
\hline
\multicolumn{3}{c}{\Rc\ filter}\\
\hline
01 & 8:18 &12& 316$\pm$6&  0.0524$\pm$0.0015& 983$\pm$37 &0.0642$\pm$0.0030 \\
02 &  8:25 &  114& 367$\pm$4&  0.2042$\pm$0.0032 &  318$\pm$6 
&0.0929$\pm$0.0023 \\
03 &  8:53 &  223 & 356$\pm$4&  0.1641$\pm$0.0020 & 1648$\pm$28 &
0.3529$\pm$0.0060 \\
04 &  8:04 &  307 & 193$\pm$4&  0.0697$\pm$0.0016&  1555$\pm$48 
&0.2023$\pm$0.0066 \\
05 &  7:08  &  28 & 108$\pm$6&  0.0178$\pm$0.0019&  103$\pm$9 
&0.0149$\pm$0.0030 \\
05  & 8:19  &  51 & 148$\pm$1&  0.0756$\pm$0.0011 & 230$\pm$3 &
0.0563$\pm$0.0014 \\
06 &  5:03  &  88 & 186$\pm$1&  0.0920$\pm$0.0011 & 126$\pm$4 &
0.0406$\pm$0.0033 \\
06 &  5:27 &   96 & 184$\pm$1&  0.0978$\pm$0.0013&  99$\pm$4 
&0.0452$\pm$0.0053 \\
06 &  7:01 &  126 & 246$\pm$1&  0.1837$\pm$0.0009 & 181$\pm$4
&0.0535$\pm$0.0023 \\
06  & 8:12  & 148 & 348$\pm$1&  0.2130$\pm$0.0010 & 274$\pm$3 
&0.0845$\pm$0.0015 \\
\hline
\end{tabular}
\end{center}
\end{table*}

\subsubsection{PA $\sim$ 90\deg}

For the activity cloud released at PA$\sim$90\deg\ the variations of $SA$
for \Bc\ and
\Rc\ filters with the phase have similar behaviors, with the
values for \Bc\ lower than those of \Rc. By
interpolating the points obtained with different phases, we see that,
multiplying the \Bc\ intensities by 1.36, the two profiles overlap. As
long as the grains are bigger than the wavelength, this is simply the
variation of the albedo from \Bc\ to \Rc, \ie\ the color of the
grains.  Note that in such a way those measurements of color are
independent from any temporal change due to the variability of the
emission. The error in the color of the grains has been evaluated as the
standard
deviation of the data with respect to an interpolating curve. It results of
the order of 0.026 km$^2$.

The albedo of the grains in \Rc\ is then 36\% higher than in
\Bc. Assuming a linear dependence with wavelength, the albedo at
$\lambda = 5550$~\AA\ is 20\% higher than that at \Bc\ and the
spectral slope is (14.7$\pm$1.1)~\%/1000~\AA.  This value is more than 3
times larger than that of the background dust emission of the ``quiet
comet'' obtained in the previous section.

In Fig.~\ref{fig_SA_phase} (a), the $SA$ derived from the
observations with the \Bc\ filters have been multiplied by 1.36 to
take into account the different albedos, in order to check how the
cross-sections change with the rotation phase. The points from
consecutive nights interleave nicely, suggesting the repeatability
of the emission period after period. However, 
we cannot firmly prove this asserting. In fact, for example, the data at
phase
around 200\deg\ come from observations of Nov. 3 at UT 8-9, while the data for
phase of
300\deg\ come from observations taken the day after, more or less at
the same UT (see tables
\ref{Bc_ejecta}). That means that the nucleus had rotated by one full turn
plus 100\deg.

Assuming a typical grain albedo of $A=0.04$, the maximum area
covered by them is of the order of 5 km$^2$. The variations of $SA$
illustrated in Fig.~\ref{fig_SA_phase} are very large: it varies from
almost 0.04~km$^2$, at phase close to 0, to more than 0.2~km$^2$ at phase
$\sim 150$--200\deg, with an increase of at least a factor of 5.  The
increases of $SA$ between phase 0\deg\ to 200\deg\ can be simply
explained with an increase of the production in the grains as the
active area moves into sunlight.

Instead the factor of four decrease passing from phase 200\deg\ to phase 300\deg
(see
Fig.~\ref{fig_SA_phase}, a),
corresponding to about 5~h if the periodicity can be trusted (or 24~h
actual elapsed time), is more difficult to explain: the cloud must
still be well within the FoV of the observations because of the low
projected velocity. This strongly supports the hypothesis that grains
sublimated away, with a lifetime surely shorter than 24~h and probably of the
order of 5~h. If they are transformed into gas, they will contribute no longer
to the observations in the continuum bands.

The color of the grains ejected at this PA is similar to that
of the surface of short period comets, Centaurs and scattered disk
TNOs \citep[see online MBOSS color database][]{HD02}.  This color
cannot be explained by pure ice; they could be ice particles with a
large amount of  organic material or just organic grains
that sublimate under the solar radiation.

\subsubsection{PA $\sim$ 140\deg}

The clouds of particles released at PA $\sim$ 140\deg\ are much fainter
and the data
therefore are more noisy than the other ones. However the data taken through the
\Bc\ filter overlap with those through the \Rc\ filter, within the errors. The
estimated error is here
about 0.05 km$^2$. The corresponding spectral slope is then 
0$\pm$2.4~\%/1000~\AA. 

This cloud appears during a short range of rotation period, at a
phase
$\sim$~200\deg.  The $SA$ changes by a factor 6--7 during the whole
rotation. As for the other cloud, it decreases by a factor 4 passing
from rotation phase of 300\deg to (360+)20\deg,  suggesting
sublimating grains for this cloud as well.

However, these grains seem to have a gray color, in contrast to those
released at PA$\sim$90\deg. The appearance of the two
clouds at different
rotation phases indicates that they are produced by different active
regions of the nucleus, and the great difference in their color, suggest that
these
regions produce different kind of grains. Those at PA$\sim$140\deg\ are more
similar to the icy chunks discovered by the spacecraft,  because
ice should
have gray color. Also those grains cannot be pure ices, since
their lifetimes should be much longer that the $\simeq$ 5 hours
indicated here (see for example \cite{Beer2006}). So also these grains should
have some impurities of gray color, as for example silicates.


\section{Conclusions}

From the analysis of images recorded  in the blue and red continuum regions of
the visible spectra, we have separated the solid state emission produced by
the periodic activity from the normal dust cometary coma of the ``quiet''
comet. 

We have found that the coma of the ``quiet'' comet shows a peculiar
behavior: its \SAf\ function is not constant, but shows an increase
with nucleo-centric distances $\rho$, for $\rho < $2000-2500 km, that
is not typical of a comet with constant outflow velocity and without
fragmentation or sublimation.
The \SAf\ profiles of the ``quiet coma'' obtained over the 5~nights of
observations have a very similar behavior. Of course this is true for
the observations where the emission due to the periodic activity does
not hide completely the inner coma. This kind of profile is an
indication that the scattering of the grains increases as they move
away from the nucleus. Since this behavior is the same during a long
period of time (5 nights), the only explanation is that the grains
present in the coma of the ``quiet'' comet fragment with the time,
increasing their scattering area with the nucleo-centric distance.


The color of the solid state coma of the ``quiet'' comet is red, with
a spectral slope of 4.6$\pm$1.5 \%/1000 \AA, assuming a linear
variation with wavelength.

By analyzing the grains produced by the periodic activity we have
shown that the released clouds have two privileged projected directions:
one at PA$\sim$90\deg\ and the other at $\sim$140\deg. The
profiles of the equivalent
\SAf\ function with $\rho$  of the clouds released in the two direction can be
fitted with a constant plus an exponential function.  The
cross-sections $SA$ of the two clouds vary strongly with the rotation
phase of the nucleus. They move very slowly, with projected velocities
of the order of 20 m/s, for both PAs. One of the two clouds (the
one with PA$\sim$90\deg) has a very red color, of the order of 15$\pm$3 \%/1000
\AA, while the other seems to have a gray color. Assuming a
periodic emission with the rotation, both sublimate with lifetimes of the
order of 5 h.

During the flyby, the spacecraft  has observed for the first time some
large chunks around the nucleus, that are supposed to be icy ice particles.
The gray cloud can be then constituted by
those chunks. However, to explain their relatively short lifetimes, those
particles also need to have some impurities, gray in color, as
for
example silicates. The clouds released in the other direction are very red and
sublimate as well with a lifetime of the order of few hours.  So  these grains
should have lot of  organics embedded in them, or they are pure organic grains
that sublimate.
It is important to notice that, within $\rho$ equal to 4000 km, \ie where most
of the activity takes place, the cross section of the clouds produced by the
activity is at
most 6\% of that of the quiet comet. This is of the
same order of magnitude of the 4\% contribution of the icy chunks
measured by the spacecraft \citep{AHearn2011}.
 
Once the full geometrical analysis of the jets observed by the
spacecraft becomes available, it will be interesting to connect them
with the two clouds described here, and to investigate whether their
different characteristics can be traced to different natures of the
sublimating area.

\section{Acknowledgments}

Based on Observations performed at the European Southern Observatory, La
Silla, Progr. 086.C-0375. This work has been partially funded by MIUR -
Ministero dell'Istruzione, dell'Universit\'a e della Ricerca (Italy),
under the PRIN 2008 funding.

\section{References}

\bibliographystyle{aa}
{}
\end{document}